# Monopole Excitations of Baryons in the Nambu–Jona–Lasinio Soliton Model[†]

Abdellatif Abada, Reinhard Alkofer, Hugo Reinhardt, and Herbert Weigel[‡]

Institute for Theoretical Physics
Tübingen University
Auf der Morgenstelle 14
D-72076 Tübingen, Germany

## ABSTRACT

We investigate the monopole excitations of the soliton in the Nambu–Jona–Lasinio model. By studying the solutions to the corresponding Bethe–Salpeter equation in the background of the soliton we exclude the existence of real large amplitude fluctuations. This allows us to treat the collective coordinate for the monopole excitations, which parametrizes the extension of the soliton, in the harmonic approximation. The canonical quantization of this coordinate yields a spectrum which agrees reasonably well with the empirical data for the Roper resonance, N(1440), and the corresponding one for the Delta, $\Delta(1600)$. We also comment on going beyond the harmonic approximation.

---

[†] Supported by the Deutsche Forschungsgemeinschaft (DFG) under contract Re 856/2-2.
[‡] Supported by a Habilitanden–scholarship of the DFG.

*1. Introduction*

Bosonization of the Nambu–Jona–Lasinio (NJL) model [1] provides an effective meson theory which predicts the properties of the pseudoscalar and vector mesons reasonably well [2]. This result has motivated its consideration as a microscopic model for the quark flavor dynamics of QCD. Over the past few years it has in addition become evident that the soliton approach to the bosonized NJL model suitably describes the properties of the low–lying $\frac{1}{2}^+$ and $\frac{3}{2}^+$ baryons [3]. In the present article we will address the question in how far this model also permits the description of excited baryons. The natural extension is to consider monopole, *i.e.*, radial excitations of the ground states. Such analyses provide descriptions of the Roper (1440) and $\Delta$ (1600) resonances, which are excited nucleon and $\Delta$ states, respectively.

Three different methods have been used to investigate the monopole channel within the related (although much simpler) Skyrme model [4, 5]: the scaling approach [6], $\pi$-$N$ phase shifts analysis [7], and the linear response theory [8]. In the phase shifts analysis [7] the Roper resonance cannot be observed because there is a almost complete cancellation between the monopole and rotational channels in the geometrical coupling scheme of ref. [10]. In the linear response theory a resonance is observed for the breathing mode at approximately 400MeV [9]. Unfortunately this resonance cannot immediately be identified with the Roper resonance because the coupling to the rotational channel was omitted in that calculation. These two approaches appear to suffer from the rotational channel not being treated as a large amplitude fluctuation, *i.e.*, like a zero mode. On the other hand the Roper resonance is clearly identified in the scaling method, which allows for a dynamical coupling between the monopole and rotational degrees of freedom. In this approach the excitation energy of this mode comes out at the order of 300MeV [6] which is somewhat too small as compared to the experimental value of 500 MeV but nevertheless considerably closer to that value than the prediction of the non–relativistic quark model [11].

In this paper we will therefore study the scaling method in the framework of the NJL model as a first approach to describe excited (non–strange) baryons. We should remark that treatments like the phase shift analysis, which involve meson excitations at arbitrary frequencies, would be plagued by the non–confining character of the NJL model. Once the frequency exceeds the binding energy of the valence quarks this quark gets scattered into the continuum. For commonly adopted parameters of the NJL model the Roper resonance lies above this threshold. In the scaling method a collective coordinate is introduced which parametrizes the extension of the soliton in addition to those which describe the large amplitude motion of the rotational zero mode. The spectrum is then obtained by canonical quantization of these coordinates. From the solution to the Bethe–Salpeter equation for monopole excitations of the soliton we will argue that an harmonic approximation for the scaling variable is indeed justified. In this approximation the feature is circumvented that the NJL soliton is non–topological, *i.e.*, by shrinking to zero size the winding number zero sector can continuously be reached while the baryon number is carried by three non–interacting valence quarks.

The present paper is organized as follows. In section 2 we recall the main issues of the bosonized NJL model and its soliton solution. In section 3 the solutions to the Bethe–Salpeter equation for monopole fluctuations off the soliton are discussed. In section 4 the scaling collective Hamiltonian is determined and quantized. As indicated above the excitation energy for the scaling mode is obtained within the harmonic approximation. The numerical results as well as related discussions are given in section 5. In section 6 we conclude and comment on going beyond the harmonic approximation. The explicit expressions for the kernel of the Bethe–Salpeter equation as well as the inertia parameter for the scaling mode are given in appendices.

## 2. The bosonized NJL model

The Lagrangian for the NJL model with scalar and pseudoscalar degrees of freedom is given as the sum of the free Dirac Lagrangian and a chirally invariant four quark interaction [1]

$$\mathcal{L} = \bar{q}(i\slashed{\partial} - \hat{m}^0)q + 2G \sum_{a=0}^{N_f^2-1} \left( (\bar{q}\frac{\lambda^a}{2}q)^2 + (\bar{q}\frac{\lambda^a}{2}i\gamma_5 q)^2 \right) \tag{2.1}$$

where $q$ denotes the quark spinor and $G$ is a dimensionful effective coupling constant. $\hat{m}^0$ is the current quark mass matrix while the matrices $\lambda^a/2$ represent the generators of the flavor group $U(N_f)$. In this paper, we will consider the case of two flavors, $N_f = 2$, and assume the isospin limit $m_u^0 = m_d^0 \equiv m^0$. Using path integral techniques the model (2.1) can be bosonized and expressed in terms of composite meson fields [2]. In Euclidean space with the Euclidean time $\tau = it$ treated as a real number, the resulting effective action is given by

$$\mathcal{A} = \mathcal{A}_m + \mathcal{A}_f. \tag{2.2}$$

Here $\mathcal{A}_m$ and $\mathcal{A}_f$ are the mesonic mass and fermionic loop contributions, respectively,

$$\begin{aligned}\mathcal{A}_m &= -\frac{1}{4G} \int d^4x \, \text{tr} \, (M^+M - \hat{m}_0(M + M^\dagger) + \hat{m}_0^2), \\ \mathcal{A}_f &= \text{Tr}\log(i\slashed{D}_E) = \text{Tr}\log(i\slashed{\partial}_E - (P_R M + P_L M^+)).\end{aligned} \tag{2.3}$$

In eq. (2.3) $P_L = (1 - \gamma^5)/2$ and $P_R = (1 + \gamma^5)/2$ are the usual helicity projectors, while "Tr" denotes the functional trace including the traces over color, flavor and Dirac indices. Furthermore, $M$ is a complex matrix which contains the scalar and pseudoscalar meson fields, $M = S + iP$. In this work we will neglect fluctuations of the scalar meson field and keep it fixed at its vacuum expectation value. In order to determine the minimum of the classical energy we will furthermore assume the hedgehog ansatz for the chiral field $U$. Hence the complex field $M$ is given by

$$M(\boldsymbol{r}) = mU_\text{H}(\boldsymbol{r}) = m\exp\left(i\boldsymbol{\tau}\cdot\hat{\boldsymbol{r}}\Theta(r)\right), \tag{2.4}$$

where $m = \langle S \rangle$ is the constituent quark mass. Demanding the pion decay constant $f_\pi = 93\text{MeV}$ and mass $m_\pi = 135\text{MeV}$ yields the parameters of the NJL model as functions of the constituent quark mass [3]. This is a consequence of the gap equation, which determines the vacuum expectation value of the scalar fields $\langle S \rangle$ in the baryon number zero sector of the NJL model. We may thus consider $m$ as the only free parameter of the model.

Substituting the hedgehog ansatz (2.4) into the expression for the mesonic part $\mathcal{A}_m$ (2.3) of the effective action yields a contribution to the classical energy (subtraction of the reference case $U = 1$ is understood)

$$E_m = m_\pi^2 f_\pi^2 \int d^3r \, (1 - \cos\Theta(r)) \,. \tag{2.5}$$

Here we have made use of the relation $G = m^0 m / m_\pi^2 f_\pi^2$ obtained from the Bethe–Salpeter equation of the bosonized NJL model [2]. In the two flavor model only the real part of the fermion determinant (2.3) $\mathcal{A}_R = \frac{1}{2}\mathrm{Tr}\log(\not{D}_E^\dagger \not{D}_E)$ differs from zero. As $\mathcal{A}_R$ is ultraviolet divergent it must be regularized. We will use Schwinger's proper time regularization [12] which introduces an O(4)–invariant cutoff $\Lambda$

$$\mathcal{A}_R = -\frac{1}{2}\int_{1/\Lambda^2}^\infty \frac{ds}{s}\mathrm{Tr}\exp(-s\not{D}_E^\dagger \not{D}_E). \tag{2.6}$$

For the static meson configurations the energy associated with (2.6) splits into valence quark and vacuum contributions [13]. Namely,

$$E_{\mathrm{val}} = \eta_{\mathrm{val}}|\epsilon_{\mathrm{val}}| \tag{2.7}$$

$$E_{\mathrm{vac}} = -\frac{1}{2}\sum_\mu \int_{1/\Lambda^2}^\infty \frac{ds}{\sqrt{4\pi s^3}} e^{-s\epsilon_\mu^2} \,. \tag{2.8}$$

Here the $\epsilon_\mu$ refer to the eigenstates of the Dirac Hamiltonian

$$h = \boldsymbol{\alpha}\cdot\boldsymbol{p} + m\beta\left(\cos\Theta(r) + i\gamma_5\boldsymbol{\tau}\cdot\hat{\boldsymbol{r}}\sin\Theta(r)\right), \tag{2.9}$$

which commutes with the grand spin operator $\boldsymbol{j} + \boldsymbol{\tau}/2$. Furthermore, $\eta_{\mathrm{val}} = 0, 1$ denotes the occupation number of the valence quark, which is the state with the lowest eigenenergy (in absolute value). This occupation number has to be adjusted to guarantee unit baryon number, i.e., $1 = \eta_{\mathrm{val}} - (1/2)\sum_\mu \mathrm{sgn}(\epsilon_\mu)$.

The self–consistent chiral angle $\Theta_{\mathrm{s.c.}}(r)$ is determined by extremizing the total energy functional[1] [14]

$$E[\Theta] = E_{\mathrm{val}} + E_{\mathrm{vac}} + E_{\mathrm{m}}. \tag{2.10}$$

In the NJL model it turns out that $E[\Theta]$ depends only moderately on the extension of the meson configuration[2]. Whether this gives rise to a large amplitude fluctuation will be subject of the next section.

## 3. Monopole fluctuations

In this section we will study the question whether or not the insensitivity of the classical energy functional (2.10) with respect to scaling variations of the self–consistent soliton causes the existence of a zero–mode type state in the monopole channel. Such a state would give rise to large amplitude fluctuations like *e.g.* isospin rotations.

---

[1] For details of the numerical procedure see ref. [15].
[2] From figure 6.1 one observes *e.g.* that $E$ changes by less than 5% when the extension is altered by 30%.

As all our computations are subject to grand spin symmetry the corresponding time dependent meson fluctuation in the monopole channel is parametrized by [7, 16]

$$\eta(\boldsymbol{r},t) = \boldsymbol{\tau} \cdot \hat{\boldsymbol{r}} \eta(r,t). \tag{3.1}$$

Obviously this *ansatz* describes a pseudoscalar P–wave pion. The expression for the chiral field, which contains both the soliton and the monopole fluctuation, reads

$$U(\boldsymbol{r},t) = \exp\left\{i\boldsymbol{\tau} \cdot \hat{\boldsymbol{r}} \left(\Theta_{\text{s.c.}}(r) + \eta(r,t)\right)\right\}. \tag{3.2}$$

The NJL–model action is next expanded up to quadratic order in the fluctuation $\eta(r,t)$. No linear term appears because the chiral angle $\Theta_{\text{s.c.}}(r)$ minimizes the static energy functional. As shown in ref. [17] the quadratic term $\mathcal{A}^{(2)}$ introduces local ($\mathcal{K}_1(r)$) and bilocal ($\mathcal{K}_2(\omega;r,r')$) kernels in Fourier space

$$\begin{aligned}\mathcal{A}^{(2)} &= \int \frac{d\omega}{2\pi} \Big\{ \int dr\, r^2 \int dr'\, r'^2 \tilde{\eta}(r,\omega) \mathcal{K}_2(\omega;r,r') \tilde{\eta}(r',-\omega) \\ &\quad + \int dr\, r^2 \mathcal{K}_1(r) \tilde{\eta}(r,\omega) \tilde{\eta}(r,-\omega) \Big\}. \end{aligned} \tag{3.3}$$

Here $\tilde{\eta}(r,\omega)$ denotes the Fourier transform of $\eta(r,t)$, *i.e.*, $\eta(r,t) = \int \frac{d\omega}{2\pi} \tilde{\eta}(r,\omega) e^{-i\omega t}$. The explicit expressions for the kernels of the monopole fluctuation in terms of the eigenvalues and –functions of the static Dirac Hamiltonian (2.9) are displayed in appendix A. Here we only wish to make a few remarks on the boundary problem. The eigenstates of (2.9) are discretized by demanding the upper component of the Dirac spinors $\Psi_\mu$ to vanish at the boundary ($r = D$) of a spherical box[1] [18]. Eventually we consider $D \to \infty$, in practical computations this means that $D$ is significantly larger than the extension of the soliton profile $\Theta(r)$. The boundary condition for $\Psi_\mu$ transfers to the kernels. As can be observed from the explicit expressions given in appendix A the bilocal kernel $\mathcal{K}_2(\omega;r,r')$ vanishes whenever either $r$ or $r'$ equals $D$ while $\mathcal{K}_1(r = D) \neq 0$. Hence the solutions to the Bethe–Salpeter equation[2]

$$r^2 \left\{ \int dr'\, r'^2 \mathcal{K}_2(\omega;r,r') \tilde{\eta}(r',\omega) + \mathcal{K}_1(r) \tilde{\eta}(r,\omega) \right\} = 0 \tag{3.4}$$

obey the boundary condition $\tilde{\eta}(D,\omega) = 0$. A well defined chiral field $U$ also requires $\tilde{\eta}(0,\omega) = 0$.

In Fig. 3.1 typical solutions to the Bethe–Salpeter equation (3.4) are shown. In case there is no soliton present our solutions are (except of a small vicinity of $r = D$) identical to spherical Bessel functions associated with unit orbital angular momentum[3]. Of course, this is just what we expect from a free P–wave fluctuation.

---

[1] An alternative set of boundary conditions is given in ref. [19].
[2] In the two flavor case $\mathcal{K}_2(\omega;r,r')$ is invariant under $\omega \leftrightarrow -\omega$.
[3] This deviation can easily be understood because in the region $r \approx D$ the completeness relation $\sum_\mu \Psi(\boldsymbol{r}) \Psi^\dagger(\boldsymbol{r}') \sim \delta(\boldsymbol{r} - \boldsymbol{r}')$ can only approximately be fulfilled in a finite model space.

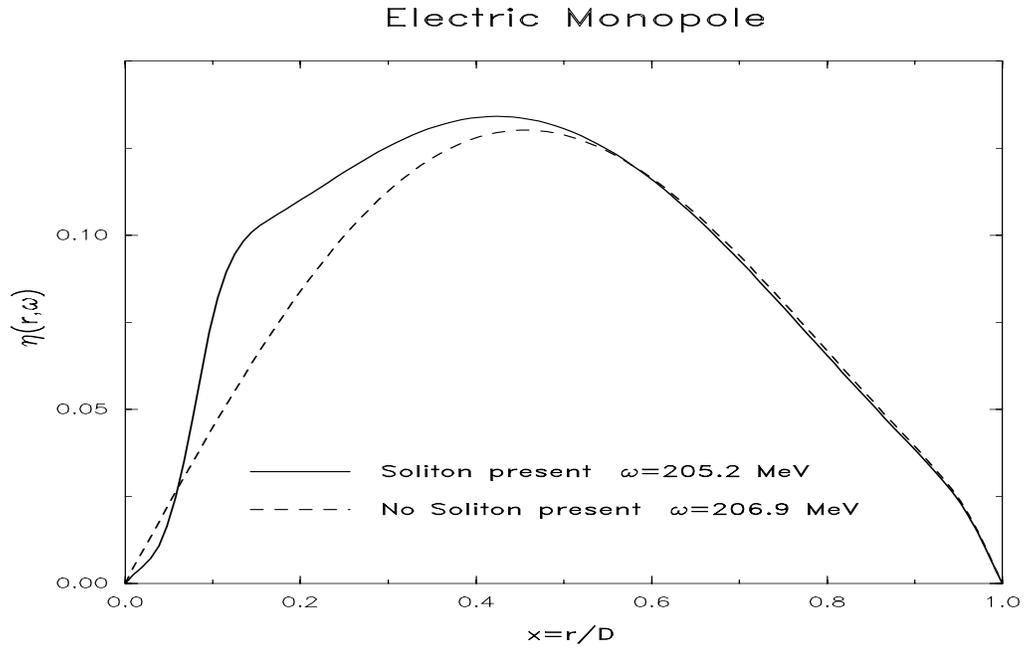

Figure 3.1: The profiles of solutions to the Bethe–Salpeter equation (3.4) for the constituent mass $m = 450$MeV. These solutions are computed using $D = 6$fm. The normalization is chosen arbitrarily.

Two important remarks are in order for the solutions to the Bethe–Salpeter equation (3.4) in the background of the self–consistent soliton. First we recognize that the corresponding solution to (3.4) deviates from the free P–wave only in the region where the soliton profile is non–trivial ($r \leq 1$fm). Secondly, and more importantly, the resulting energies of the eigenmodes in the presence of the soliton agree with those of free P–wave solutions at the order of 1%. Usually the existence of large amplitude fluctuations causes a strong reduction of the eigenenergy, it may even vanish in the case that the fluctuation corresponds to an exact symmetry. We therefore conclude that no large amplitude fluctuation exists in the monopole channel despite of the insensitivity of the classical energy with respect to scaling variations. Moreover, the spectrum resulting from the quantization of the collective scaling variable will be dominated by the properties of the corresponding Hamiltonian in the vicinity of the value which minimizes the potential. These considerations justify the harmonic approximation for this collective mode which will be discussed in the proceeding section.

## 4. The scaling collective Hamiltonian

The collective breathing mode of the soliton is described by the time–dependent coordinate $\chi(t)$ which parametrizes the extension of the meson configuration, $\Theta_\chi(r,t) = \Theta_{\mathrm{s.c.}}(\chi(t)r)$. States with good spin and isospin quantum numbers are generated within the cranking procedure which requires collective coordinates $R(t) \in \mathrm{SU}(2)$ for the (iso) rotations. We therefore consider the meson configuration

$$M(\boldsymbol{r},t) = m R^\dagger(t) U_{\mathrm{H}}(\chi(t)\boldsymbol{r}) R(t) \tag{4.1}$$

where $U_{\mathrm{H}}$ is the hedgehog soliton (2.4). This configuration is substituted into the regularized NJL model action and an expansion up to second order in the time derivative is performed. From this the collective Lagrangian [6]

$$L(\chi,\dot\chi) = -E(\chi) + \frac{1}{2}a(\chi)\dot\chi^2 + \frac{1}{2}\alpha^2(\chi)\boldsymbol{\Omega}^2 \tag{4.2}$$

is extracted. $E(\chi) = E[\Theta_\chi]$ is the energy functional defined in section 2, however, evaluated using $\Theta_{\mathrm{s.c.}}(\chi(t)r)$ at a fixed time. I.e., we have substituted the eigenvalues of

$$h_\chi = \boldsymbol{\alpha}\cdot\boldsymbol{p} + m\beta\left(\cos\Theta_\chi + i\gamma_5\boldsymbol{\tau}\cdot\hat{\boldsymbol{r}}\sin\Theta_\chi\right) \tag{4.3}$$

into eqs. (2.7) and (2.8). The mesonic part of the action (2.3) only contributes to $E(\chi)$. The inertia parameter $a(\chi)$ for the scaling mode may be interpreted as a position dependent mass for the collective coordinate $\chi$. Its explicit form in terms of the eigenvalues and –functions of $h_\chi$ is displayed in appendix B. Furthermore $\boldsymbol{\Omega}$ is the time derivative of the collective rotations $R^\dagger \dot R = (i/2)\boldsymbol{\tau}\cdot\boldsymbol{\Omega}$. The position dependent moment of inertia $\alpha^2(\chi)$ is obtained from the expression given in the literature [13, 3] by again substituting the eigenvalues and –functions of $h_\chi$ (cf. appendix B).

In order to avoid ordering ambiguities in the process of quantization we perform a variable transformation $\chi = \chi(\xi)$ such that $a(\chi(\xi))[d\chi/d\xi]^2 = 1$. This transformation is allowed as long as $a(\chi) > 0$ which actually is found to hold in our numerical studies. It is then straightforward to obtain the collective Hamiltonian in terms of the coordinate $\xi$ and its conjugate momentum $\hat{P} = -id/d\xi$

$$H_J = \frac{1}{2}\hat{P}^2 + \tilde{V}_J(\xi) \tag{4.4}$$

where

$$\tilde{V}_J(\xi) = V_J(\chi(\xi)) \quad \text{with} \quad V_J(\chi) = E(\chi) + \frac{J(J+1)}{2\alpha^2(\chi)}. \tag{4.5}$$

We have already inserted the eigenvalue $J$ of the spin operator, which represents the momentum conjugate to $R$. This procedure is justified because the generators for rotations and scalings commute.

The above motivated harmonic expansion is carried out separately in each spin channel. For this purpose the minimum of the potential $\tilde{V}_J(\xi)$ is determined. We denote its position by $\xi_J^{\min}$. Then the spectrum of the collective Hamiltonian (4.4) is approximated by

$$E_{J,n} = \tilde{V}_J(\xi_J^{\min}) + \omega_J\left(n + \frac{1}{2}\right), \quad \omega_J = \sqrt{\left.\frac{\partial^2 \tilde{V}_J(\xi)}{\partial \xi^2}\right|_{\xi_J^{\min}}} = \sqrt{\frac{V_J''(\chi_J^{\min})}{a(\chi_J^{\min})}} \tag{4.6}$$

with $\chi_J^{\min} = \chi(\xi_J^{\min})$ and the prime indicating a derivative with respect to $\chi$. It should be remarked that $\chi_J^{\min}$ minimizes $V_J(\chi)$. The harmonic approximation has obviously the advantage that we do not have to explicitly carry out the change of variables $\chi \to \xi$. A further justification of this approximation is provided by the fact that our computation of $\alpha^2(\chi)$ cannot be generalized to arbitrary large $\chi$, cf. the discussion at the end of appendix B.

The nucleon corresponds to the quantum numbers $J = 1/2$ and $n = 0$ while the Roper resonance is associated with $J = 1/2$ and $n = 1$. As an illustrative example we also list the expression for the $\Delta$-nucleon mass difference

$$M_\Delta - M_N = V_{3/2}(\chi_{3/2}^{\min}) - V_{1/2}(\chi_{1/2}^{\min}) + \frac{1}{2}\left(\omega_{3/2} - \omega_{1/2}\right). \tag{4.7}$$

Obviously this mass difference acquires an additional contribution, which is due the treatment of the scaling mode as a quantum variable.

## 5. Numerical results

In this section we present our predictions obtained in the NJL soliton model for the masses of excited baryons in the monopole channel using the harmonic approximation to the breathing mode. It is known that the absolute masses of solitons acquire substantial reductions associated with meson loop corrections [20]. Although the absolute mass of the soliton in the NJL model is not as large as in the Skyrme model (for parameters fitted to the meson sector) these reductions are relevant in the NJL model as well [21]. As they effect all baryon states approximately equally we will concentrate on mass differences only.

As already mentioned above the constituent quark mass $m$ is the only undetermined parameter. Although for a unit baryon number soliton solutions exist for $m \geq 325$MeV [14] these solutions only represent local minima of the energy $E$ (2.10). Unless $m \geq 420$MeV the configuration consisting of three non–interacting valence quarks is energetically favored against the soliton configuration. As the NJL soliton is of non–topological character it can continuously be deformed from the local to the global minimum by shrinking it to zero size ($\chi \to \infty$) without encountering an infinite energy barrier. We therefore consider only the region in parameter space where the soliton represents the global minimum of the energy functional *i.e.*, $m \geq 450$MeV. A value of $m$ that large is also mandatory to find a pronounced minimum of $V_J$ (4.5) in the $\Delta$ channel ($J = 3/2$).

In table 5.1 the numerical results for quantities appearing in the mass formula (4.6) are shown for various values of $m$. We have first computed the position $\chi_J^{\min}$ of the minimum of $V_J(\chi)$ (4.5). As the moment of inertia cancels from eq. (4.6) for $J = 0$ the result $\chi_0^{\min} = 1$ confirms that $\Theta_{\rm s.c.}(r)$ indeed minimizes the classical energy. We observe that the deviation of $\chi_J^{\min}$ from unity is significantly smaller in the NJL model than the Skyrme model calculations [6] which yield $\chi_{3/2}^{\min}$ as small as 0.4. Although the breathing mode potential is shallow it is at least steeper than in the Skyrme model. Subsequently $\chi_J^{\min}$ has been employed to evaluate the classical mass $E$, the moment of inertia $\alpha^2$ and the breathing frequency $\omega_J$ defined in eq. (4.6). Obviously $\omega_{1/2}$ represents our prediction for the mass difference between the Roper resonance and the nucleon. From table 5.1 we also deduce that the classical energy $E$ is indeed quite insensitive to variations in the scaling variable. Glancing *e.g.* at the case $m=500$MeV shows that $E$ changes by only about 5% when $\chi$ is reduced by 20%. On the other hand the moment of inertia $\alpha^2$ and the breathing frequency $\omega_J$ crucially depend on $\chi$.

Now we come to the central issue of this paper, the spectrum of the scaling mode. If this mode were not treated quantum mechanically the $\Delta$ nucleon mass difference would be equal $3/2\alpha^2$. The experimental value of this mass difference (293MeV) corresponds to $\alpha^2 \approx 5.12\text{GeV}^{-1}$ which is obtained for $m \approx 420$MeV. However, as already indicated at the end of section 4 the treatment of the scaling mode as a quantum variable drastically alters this result. From table 5.2 we observe that this mass difference is best reproduced for values of the constituent quark mass as large as $m \approx 550$MeV. This result also gives an *a posteriori* justification for the harmonic approximation which requires a pronounced minimum of $V_J$ (4.5). This is not the case for small constituent quark masses.

We also find that the parameter $m \approx 550$MeV not only correctly reproduces the $\Delta$–nucleon splitting but also leads to a reasonable prediction for the mass difference between the Roper resonance and the nucleon. As a matter of fact the Roper nucleon splitting is almost independent of the constituent quark mass. It appears to be a common feature of the scaling approach to soliton models that the Roper comes out on the low side [6]. For $m = 550$MeV the agreement for the first excitation above the $\Delta$ is equally good although its position is more dependent on $m$. In table 5.2 we have also displayed our predictions for second excited states in the $J = 1/2$ and $J = 3/2$ channels. Also in these cases our results compare reasonably well with the experimental data although these are not exactly identified[1].

---

[1] In ref. [22] these states are listed as three star resonances.

Table 5.1: The parameters of mass formula (4.6) at $\chi_J^{\min}$ which minimizes the potential $V_J(\chi)$ (4.5).

| | | $m$=450MeV | | |
|---|---|---|---|---|
| $J$ | $\chi_J^{\min}$ | $E$(GeV) | $\alpha^2$(1/GeV) | $\omega_J$(GeV) |
| 0 | 1.00 | 1.232 | 4.79 | 0.450 |
| 1/2 | 0.99 | 1.233 | 4.80 | 0.400 |
| 3/2 | 0.83 | 1.269 | 5.47 | 0.266 |
| | | $m$=500MeV | | |
| $J$ | $\chi_J^{\min}$ | $E$(GeV) | $\alpha^2$(1/GeV) | $\omega_J$(GeV) |
| 0 | 1.00 | 1.221 | 4.17 | 0.456 |
| 1/2 | 0.97 | 1.222 | 4.24 | 0.403 |
| 3/2 | 0.79 | 1.287 | 5.46 | 0.303 |
| | | $m$=550MeV | | |
| $J$ | $\chi_J^{\min}$ | $E$(GeV) | $\alpha^2$(1/GeV) | $\omega_J$(GeV) |
| 0 | 1.00 | 1.208 | 3.75 | 0.461 |
| 1/2 | 0.96 | 1.210 | 3.89 | 0.405 |
| 3/2 | 0.77 | 1.294 | 5.46 | 0.324 |
| | | $m$=600MeV | | |
| $J$ | $\chi_J^{\min}$ | $E$(GeV) | $\alpha^2$(1/GeV) | $\omega_J$(GeV) |
| 0 | 1.00 | 1.193 | 3.46 | 0.465 |
| 1/2 | 0.95 | 1.196 | 3.65 | 0.404 |
| 3/2 | 0.75 | 1.294 | 5.45 | 0.354 |

Table 5.2: The predictions of the masses of the low–lying baryons according to the mass formula (4.6). Given are the mass differences to the nucleon $J = 1/2$, $n = 0$. The energy dimension is MeV.

| | | | $m$ | | | | |
|---|---|---|---|---|---|---|---|
| $J$ | $n$ | | 450 | 500 | 550 | 600 | Expt. [22] |
| 3/2 | 0 | $\Delta(1232)$ | 234 | 270 | 292 | 315 | 293 |
| 1/2 | 1 | $N(1440)$ | 400 | 403 | 405 | 404 | 501 |
| 3/2 | 1 | $\Delta(1600)$ | 500 | 573 | 616 | 669 | 661 |
| 1/2 | 2 | $N(1710)$ | 800 | 806 | 810 | 808 | 742 − 802 |
| 3/2 | 2 | $\Delta(1920)$ | 766 | 876 | 940 | 1023 | 962 − 1032 |

## 6. Conclusions

We have investigated the monopole excitations of baryons within the NJL chiral soliton model. Although the surface of the classical energy is quite flat in the direction of scaling variations of the static soliton it has turned out that no large amplitude fluctuation exists for the scaling mode. In account of this result we have argued that the spectrum in the monopole channel is characterized by the properties which the potential exhibits at its minimum. When elevating the scaling mode to a quantum variable we have therefore treated this potential in the harmonic approximation. Adopting the constituent quark mass $m = 550\text{MeV}$ within this approach we have obtained a reasonable agreement with the available data for the mass differences of the exited non–strange baryons in the $J = 1/2$ and $J = 3/2$ channels. Although this value for the constituent quark mass appears to be somewhat high there is nothing special about, it just represents the only free parameter of the model. On the contrary, such a large value is appreciated since it makes the harmonic approximation more reliable since the minimum of the potential is more pronounced thereby providing an *a posteriori* justification of the method. Let us nevertheless comment on treating the Hamiltonian (4.5) beyond the harmonic approximation. For this purpose we have displayed the complete dependence of the potential $V_J(\chi)$ on the scaling variable $\chi$ in figure 6.1. Obviously the potential stays finite as $\chi \to \infty$, rather $E[\Theta] \to 3m$. This limit just corresponds to the absence of the soliton (shrunk to a point) and the baryon number carried by three non–interacting valence quarks. Of course, in a confining model such a minimum would not exist. Stated otherwise, our harmonic approximation represents a (crude) way to imitate confinement. For our preferred value $m = 550\text{MeV}$ of the constituent quark mass we find that decays of the first excited states ($n = 1$) into three free quarks are on the border of being energetically forbidden. The higher excited states may, after passing through a finite energy barrier, decay. As the full potential is more shallow than the one approximated harmonically the predictions for the masses of the excited states will be reduced. It should also be mentioned that any treatment which goes beyond the harmonic approximation suffers from ordering ambiguities when quantizing the breathing coordinate especially because $a(\chi)$ may contain large derivatives.

There is an additional feature we can read off from figure 6.1. Although we observe quite a pronounced minimum of the potential for the soliton, it becomes the more shallow the larger the spin quantum number is. For $J > 3/2$ a minimum ceases to exist. This fact may be considered as an indication that the NJL soliton model does not contain baryons with spin larger than $J = 3/2$ when the rotational degrees of freedom are treated beyond the cranking approximation. Of course, this is expected [23] within a model of baryons which is formulated in terms of quark degrees of freedom when $N_C = 3$ is adopted for the number of colors.

## Appendix A: Kernels for monopole fluctuations

In this appendix we will present the explicit expressions for the kernels which enter the quadratic form for the monopole fluctuation (3.3). The expressions quoted in this appendix refer to Minkowski space.

It is suitable to introduce the chiral rotation

$$\mathcal{T}(\boldsymbol{r}) = \cos\frac{\Theta(r)}{2} - i\gamma_5 \boldsymbol{\tau} \cdot \hat{\boldsymbol{r}} \sin\frac{\Theta(r)}{2} \tag{A.1}$$

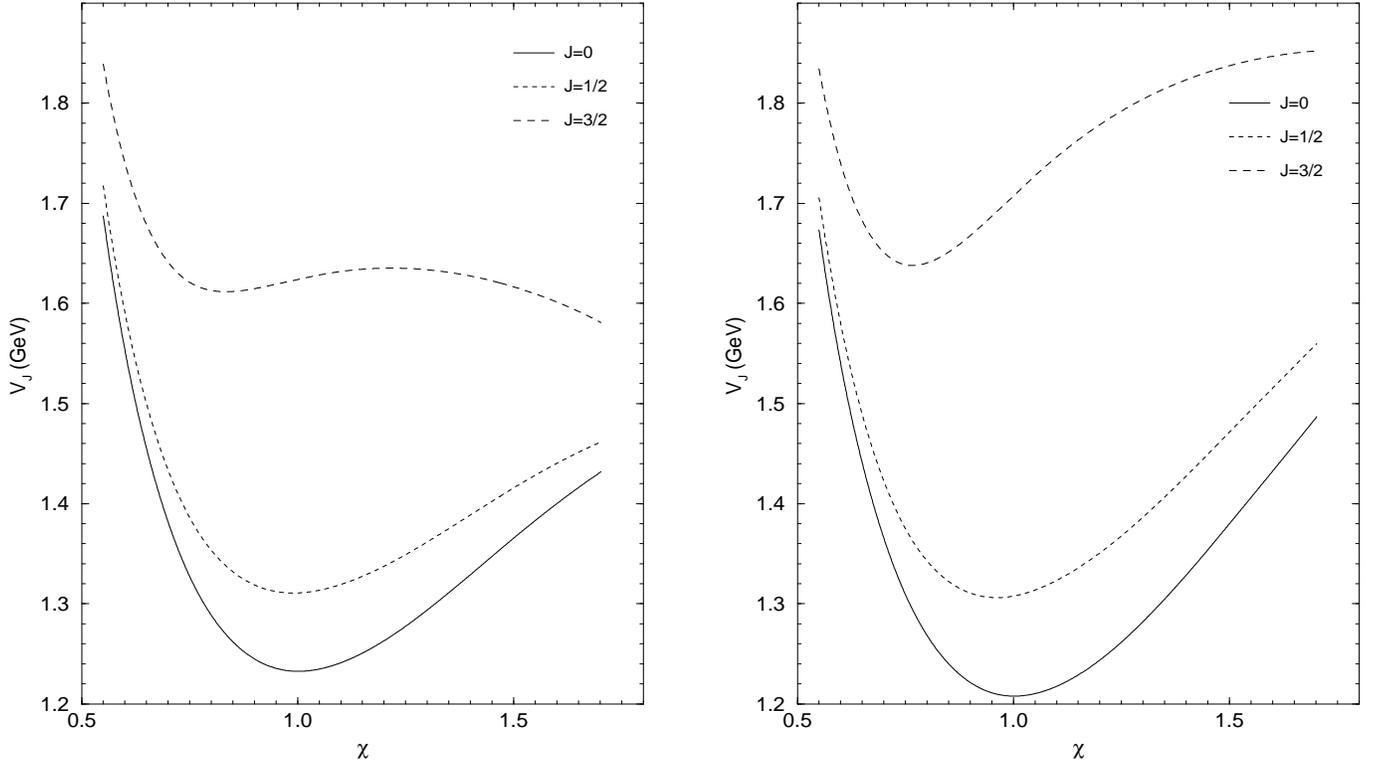

Figure 6.1: The potential of the collective Hamiltonian (4.5) for the spin quantum numbers $J = 0, 1/2$ and $3/2$ as function of the scaling coordinate $\chi$. The constituent masses adopted here are $m = 450\text{MeV}$ (left panel) and $m = 550\text{MeV}$ (right panel).

since it permits a compact notation for the Dirac operator in the presence of the monopole fluctuations (3.1)

$$i\beta\rlap{/}D = i\partial_t - \boldsymbol{\alpha} \cdot \boldsymbol{p} - m\mathcal{T}(\boldsymbol{r})\beta \left\{1 + i\gamma_5 \boldsymbol{\tau} \cdot \hat{\boldsymbol{r}}\eta(r,t) - \frac{1}{2}\eta(r,t)^2 + \ldots\right\} \mathcal{T}^{\dagger}(\boldsymbol{r}). \tag{A.2}$$

When computing the functional trace the chiral rotation can straightforwardly be imposed onto the eigenfunctions of the static Dirac Hamiltonian (2.9). This procedure simplifies the expressions for the direct coupling between the soliton and the meson fluctuation. Since the eigenfunctions of the static Dirac Hamiltonian depend on the soliton profile functionally there is also an indirect coupling.

Substitution of the expansion (A.2) into the general expressions for $\mathcal{A}^{(2)}$ given in ref. [17] yields the local kernel which also gains a contribution form the mesonic part of the action (2.3)

$$\begin{aligned}\mathcal{K}_1(r) &= -2\pi m_\pi^2 f_\pi^2 \cos\Theta(r) - mN_C\eta_{\text{val}} \int \frac{d\Omega}{4\pi} \psi_{\text{val}}^{\dagger}(\boldsymbol{r})\mathcal{T}(\boldsymbol{r})\beta\mathcal{T}^{\dagger}(\boldsymbol{r})\psi_{\text{val}}(\boldsymbol{r}) \\ &+ \frac{1}{2}mN_C\eta_{\text{val}} \sum_\mu \text{sgn}(\epsilon_\mu)\,\text{erfc}\left(\left|\frac{\epsilon_\mu}{\Lambda}\right|\right) \int \frac{d\Omega}{4\pi}\psi_\mu^{\dagger}(\boldsymbol{r})\mathcal{T}(\boldsymbol{r})\beta\mathcal{T}^{\dagger}(\boldsymbol{r})\psi_\mu(\boldsymbol{r}),\end{aligned} \tag{A.3}$$

where we have indicated the average over the angular degrees of freedom. Similarly the bilocal kernel is obtained as

$$\begin{aligned}\mathcal{K}_2(\omega;r,r') &= m^2 N_C\eta_{\text{val}} \sum_{\nu\neq\text{val}} \frac{\epsilon_\nu - \epsilon_{\text{val}}}{(\epsilon_\nu - \epsilon_{\text{val}})^2 - \omega^2} \int \frac{d\Omega}{4\pi} \int \frac{d\Omega'}{4\pi} \\ &\quad \times \psi_{\text{val}}^{\dagger}(\boldsymbol{r})\mathcal{T}(\boldsymbol{r})\beta\gamma_5\boldsymbol{\tau}\cdot\hat{\boldsymbol{r}}\mathcal{T}^{\dagger}(\boldsymbol{r})\psi_\nu(\boldsymbol{r})\psi_\nu^{\dagger}(\boldsymbol{r}')\mathcal{T}(\boldsymbol{r}')\beta\gamma_5\boldsymbol{\tau}\cdot\hat{\boldsymbol{r}}'\mathcal{T}^{\dagger}(\boldsymbol{r}')\psi_{\text{val}}(\boldsymbol{r}') \\ &+ \frac{1}{4}m^2 N_C \sum_{\mu\nu} R(\epsilon_\mu,\epsilon_\nu;\omega^2) \int \frac{d\Omega}{4\pi} \int \frac{d\Omega'}{4\pi} \\ &\quad \times \psi_\mu^{\dagger}(\boldsymbol{r})\mathcal{T}(\boldsymbol{r})\beta\gamma_5\boldsymbol{\tau}\cdot\hat{\boldsymbol{r}}\mathcal{T}^{\dagger}(\boldsymbol{r})\psi_\nu(\boldsymbol{r})\psi_\nu^{\dagger}(\boldsymbol{r}')\mathcal{T}(\boldsymbol{r}')\beta\gamma_5\boldsymbol{\tau}\cdot\hat{\boldsymbol{r}}'\mathcal{T}^{\dagger}(\boldsymbol{r}')\psi_\mu(\boldsymbol{r}').\end{aligned} \tag{A.4}$$

Both, the local as well as the bilocal kernels, are decomposed into valence ($\sim \eta_{\text{val}}$) and vacuum contributions according to eqs. (2.7,2.8). As at large $|r|$ the chiral rotation $\mathcal{T}$ equals unity upper and lower components get connected. This causes $\mathcal{K}_2(\omega;r,r')$ to vanish when either $r$ or $r'$ lies on the boundary. The regulator function in (A.4) is given by a Feynman parameter integral

$$\begin{aligned}R(\epsilon_\mu,\epsilon_\nu;\omega^2) &= \int_{1/\Lambda^2}^{\infty} ds \sqrt{\frac{s}{4\pi}} \Bigg\{\frac{e^{-s\epsilon_\mu} - e^{-s\epsilon_\nu}}{s} \\ &\quad + \left[\omega^2 - (\epsilon_\mu + \epsilon_\nu)^2\right] \int_0^1 dx \, \exp\left[-s\left(x\epsilon_\mu^2 + (1-x)\epsilon_\nu^2 - x(1-x)\omega^2\right)\right]\Bigg\}\end{aligned} \tag{A.5}$$

which describes the quark loop in the background of the static soliton.

Although we have left the number of colors $N_C$ as a free parameter, it is always implied that it assumes its physical value $N_C = 3$.

## Appendix B: Inertia parameters

In order to extract the collective mass $a(\chi)$ for the scaling variable $\chi(t)$ within the NJL model we parametrize the time dependence of $\chi(t)$ as

$$\chi(t) = \chi_0 + \delta(t) \tag{B.1}$$

and expand the action up to quadratic order in $\delta(t)$ while keeping all orders in $\chi_0$, which is assumed to be time independent. There are various ways to perform this computation. One might *e.g.* straightforwardly adopt the treatment of ref. [17] where time dependent fluctuations off the chiral soliton have been considered. In the present case a more elegant way is to define the translation operator

$$\hat{T} = \exp\left(\delta(t)\partial_{\chi_0}\right). \tag{B.2}$$

Then the meson configuration (4.1) may be written as

$$M(\boldsymbol{r},t) = mR^\dagger(t)\hat{T}^{-1}U_{\rm H}(\chi_0\boldsymbol{r})\hat{T}R(t). \tag{B.3}$$

This parametrization may be transferred to the Dirac operator (2.3) defining the Dirac Hamiltonian

$$i\beta\rlap{/}{D} = \hat{T}^{-1}R^\dagger\left(i\partial_t - h_\chi\right)R\hat{T} \quad \text{with} \quad h_\chi = h_{\chi_0} + \frac{1}{2}\boldsymbol{\tau}\cdot\boldsymbol{\Omega} + i\dot{\delta}\partial_{\chi_0}. \tag{B.4}$$

In addition to the Coriolis term an expression involving the time derivative of the scaling coordinate has been induced, $\dot{\delta} = \dot{\chi}$. Since the eigenstates of $h_{\chi_0}$ are properly normalized the translation operator $\hat{T}$ is actually unitary. This allows us to absorb both operators $\hat{T}$ and $R$ by redefining the quark fields $q' = \hat{T}Rq$. The collective mass $a(\chi_0)$ is then extracted from the term, which is proportional to $\dot{\delta}^2 = \dot{\chi}^2$, of the NJL model action. This computation is completely analogous to standard determination of the moment of inertia $\alpha^2(\chi_0)$ [13] resulting in

$$a(\chi_0) = a_{\rm val}(\chi_0) + a_{\rm vac}(\chi_0) \tag{B.5}$$

with

$$\begin{aligned} a_{\rm val}(\chi_0) &= 2\eta_{\rm val}N_C \sum_{\mu\neq{\rm val}} \frac{|\langle\mu|\partial_{\chi_0}|{\rm val}\rangle|^2}{\epsilon_\mu - \epsilon_{\rm val}} \\ a_{\rm vac}(\chi_0) &= N_C \sum_{\mu\nu} |\langle\mu|\partial_{\chi_0}|\nu\rangle|^2 f_\Theta(\epsilon_\mu,\epsilon_\nu;\Lambda) \end{aligned} \tag{B.6}$$

with the cut-off function

$$f_\Theta(\epsilon_\mu, \epsilon_\nu; \Lambda) = \frac{\Lambda}{\sqrt{\pi}} \frac{e^{-(\epsilon_\mu/\Lambda)^2} - e^{-(\epsilon_\nu/\Lambda)^2}}{\epsilon_\nu^2 - \epsilon_\mu^2} - \frac{\text{sgn}(\epsilon_\nu)\text{erfc}\left(\left|\frac{\epsilon_\nu}{\Lambda}\right|\right) - \text{sgn}(\epsilon_\mu)\text{erfc}\left(\left|\frac{\epsilon_\mu}{\Lambda}\right|\right)}{2(\epsilon_\mu - \epsilon_\nu)}. \tag{B.7}$$

Actually $f_\Theta(\epsilon_\mu, \epsilon_\nu; \Lambda)$ is proportional to $\partial R(\epsilon_\mu, \epsilon_\nu; \omega^2)/\partial\omega^2|_{\omega=0}$. This just reflects the fact that we have treated $\dot\chi$ as a small fluctuation and expanded up quadratic order in the time derivative. It is important to note that in eq. (B.6) the eigenstates and –values of $h_{\chi_0}$ have to be substituted. In practice it is convenient to employ the identity

$$\begin{aligned}\langle\mu|\partial_{\chi_0}|\nu\rangle &= \frac{1}{\epsilon_\mu - \epsilon_\nu}\langle\mu|[h_{\chi_0}, \partial_{\chi_0}]|\nu\rangle \tag{B.8}\\ &= \frac{m}{\epsilon_\mu - \epsilon_\nu}\langle\mu|\beta r\Theta'_{\text{s.c.}}(\chi_0 r)[\sin\Theta_{\text{s.c.}}(\chi_0 r) - i\gamma_5\boldsymbol{\tau}\cdot\hat{\boldsymbol{r}}\cos\Theta_{\text{s.c.}}(\chi_0 r)]|\nu\rangle\end{aligned}$$

where the prime indicates a derivative with respect to the argument. The regulator function (B.7) vanishes for identical energies. Therefore the limit

$$\lim_{\epsilon_\mu \to \epsilon_\nu} \frac{1}{(\epsilon_\mu - \epsilon_\nu)^2} f_\Theta(\epsilon_\mu, \epsilon_\nu; \Lambda)$$

is finite.

For completeness we also give the explicit expression for the moment of inertia [13]

$$\alpha^2(\chi_0) = \alpha^2_{\text{val}}(\chi_0) + \alpha^2_{\text{vac}}(\chi_0) \tag{B.9}$$

with

$$\begin{aligned}\alpha^2_{\text{val}}(\chi_0) &= \frac{1}{2}\eta_{\text{val}} N_C \sum_{\mu\neq\text{val}} \frac{|\langle\mu|\tau_3|\text{val}\rangle|^2}{\epsilon_\mu - \epsilon_{\text{val}}} \\ \alpha^2_{\text{vac}}(\chi_0) &= \frac{1}{4} N_C \sum_{\mu\nu} |\langle\mu|\tau_3|\nu\rangle|^2 f_\Theta(\epsilon_\mu, \epsilon_\nu; \Lambda)\end{aligned} \tag{B.10}$$

where $\tau_3$ denotes a Pauli matrix. Again, $|\mu\rangle$ and $\epsilon_\mu$ refer to the eigenstates and eigenvalues of $h_{\chi_0}$, respectively. One word of caution is necessary when considering the limit $\chi_0 \to \infty$. Then the soliton actually is absent and the eigenstates of $h_{\chi_0}$ are also eigenstates of the isospin operator $\tau_3$. Since the diagonal elements of the regulator function vanish [13] $f_\Theta(\epsilon_\mu, \epsilon_\mu; \Lambda) = 0$ the moment of inertia vanishes in the limit $\chi_0 \to \infty$. Our numerical calculation does not exhibit this feature. The reason being that states employed to diagonalize $h_{\chi_0}$ are no eigenstates of $\tau_3$ rather they are eigenstates of the so–called grand spin operator, which is the sum of the total spin and isospin. The perturbation expansion of $h_\chi$ (B.4) cannot straightforwardly be generalized to $\chi_0 \to \infty$ in the grand spin basis because states with different grand spin eigenvalue (but identical spin and isopin) become degenerate. Since our techniques are unable to remove this degeneracy the small energy denominators in eq. (B.10) cause $\alpha^2$ to increase for large $\chi_0$ rather than to decrease. Similar arguments hold for $\chi_0 \to 0$ since $h_{\chi_0=0}$ is also isospin invariant. These deficiences are fortunately circumvented by the harmonic approximation (4.6).